\documentclass[%
 reprint,
superscriptaddress,
groupedaddress,
nofootinbib,
 amsmath,amssymb,
 aps,
]{revtex4-1}

\usepackage{graphicx}% Include figure files
\usepackage{dcolumn}% Align table columns on decimal point
\usepackage{bm}% bold math
\begin{document}

\preprint{APS/123-QED}

\title{Generalized tracker quintessence models for dark energy}% Force line breaks with \\
\author{L. Arturo Ure\~na-L\'opez}
 \email{lurena@ugto.mx}
\affiliation{%
 Departamento de F\'isica, DCI, Campus Le\'on, Universidad de Guanajuato, 37150, Le\'on, Guanajuato, M\'exico
}%

\author{Nandan Roy}  
 \email{royn@nu.ac.th}
 \affiliation{%
The Institute for Fundamental Study   ``The Tah Poe Academia Institute", Naresuan University, Phitsanulok 65000, Thailand}

\date{\today}% It is always \today, today,

\begin{abstract}
We study the dynamical properties of tracker quintessence models using a general parametrization of their corresponding potentials, and show that there is a general condition for the appearance of a tracker behavior at early times. Likewise, we determine the conditions under which the quintessence tracker models can also provide an accelerating expansion of the universe with an equation of state closer to $-1$. Apart from the analysis of the background dynamics, we also include linear density perturbations of the quintessence field in a consistent manner and using the same parametrization of the potential, with which we show the influence they have on some cosmological observables. The generalized tracker models are compared to observations, and we discuss their appropriateness to ameliorate the fine-tuning of initial conditions and their consistency with the accelerated expansion of the Universe at late times.
\end{abstract}

\maketitle

%\tableofcontents

\section{\label{sec:introduction}Introduction}
Dark energy (DE) is, according to our current knowledge, the main matter component of the Universe nowadays, and we also believe that is the main responsible for the accelerated cosmic expansion. The simplest possibility, a positive cosmological constant, seems to be sufficient to guarantee a good agreement of the so-called $\Lambda$CDM model with cosmological observations~\cite{Aghanim:2018eyx}.

Most studies of DE focus their attention in the so-called background evolution, as they assume that DE is a homogeneous and isotropic component whose main properties are basically encoded in its equation of state (EoS), and for that one may assume a given evolution history for it in terms of a particular parametrization. These phenomenological studies are convenient to search for deviations of the late-time cosmic expansion away from the expected one driven by a cosmological constant. But the question remains about the physical origin of such DE behavior, and then the motivation remains to search for DE parametrizations that can be connected to models with some physical interest.

Additionally, there is recent interest in studying DE density perturbations and to look for their signature in structure formation variables\cite{Abramo:2007iu,Escamilla-Rivera:2016aca,Bhattacharyya:2019lvg}, see also\cite{Perrotta:1998vf,Baccigalupi:2001aa,Amendola:2005ad}. DE is assumed to affect structure formation by changing the expansion history of the Universe\cite{Linder:2003dr,Baccigalupi:2001aa}, specially at late times. This is clearly convenient for the standard $\Lambda$CDM model, but also for other DE models as one can avoid the hassle of solving the equations of density perturbations. However, it is undeniable that the study of DE, beyond the standard cosmological constant case, must take into account its density perturbations for reliable observational constraints. 

Our aim here is to revisit quintessence DE models\cite{Amendola:2015ksp} taking into account the presence of linear density perturbations. In particular, we will consider the case of tracker quintessence, which has the nice feature to avoid an excessive fine tuning of its initial conditions. Tracker quintessence potentials were first studied in~\cite{Peebles:1987ek,Ratra:1987rm}, and the tracker condition later introduced in~\cite{Zlatev:1998yg,Steinhardt:1999nw} (see also~\cite{Ooba:2018dzf,Pavlov:2013nra} and references therein for an updated discussion on tracker potentials in flat and non-flat geometries).  For the potentials explored in the literature, it was concluded that tracker potentials were unable to provide the accelerated expansion of the Universe that is required by cosmological observations, unless their evolution behavior resembles that of the cosmological constant~\cite{Ooba:2018dzf} (which is in contrast to the early expectations on this type of models discussed in the seminal papers~\cite{Peebles:1987ek,Ratra:1987rm}).
%\textbf{As an exception the inverse-power-law potentials can drive the accelerated expansion but achieving the equation of state lower than $-0.8$ is difficult for these models \cite{Zlatev:1998yg, Steinhardt:1999nw}. This makes these potentials incompatible with the current observations in which the equation of state of the dark energy $w_{DE} = −1.03 \pm 0.03$~\cite{Aghanim:2018eyx}.}

However, it can be shown that there is a generalized class of quintessence potentials that possess the tracker characteristic and whose late-time evolution can be close to that of a cosmological constant. For this, we will use the same formalism presented in Refs.~\cite{Roy:2018nce,Urena-Lopez:2015gur,Urena-Lopez:2015odd,Cedeno:2017sou}, which considered a transformation of the quintessence equations of motion into a dynamical system by choosing properly defined variables in the polar form. The presence of the so-called active model parameters is clearly shown in the new equations of motion, and this is used to put constraints by means of comparison with observations.

The organization of the manuscript is as follows. In Sec.~\ref{sec:mathematical} we introduce the equations of motion for the evolution of background and perturbation variables, for both cases in the form of a dynamical system. In Sec.~\ref{sec:critical}, we discuss in general terms the general condition for tracker solutions using a particular parametrization of the quintessence models, although we will also discuss its applicability to other choices. In Sec.~\ref{sec:numerical}, we present the numerical studies of the quintessence models and the observational constraints upon the free parameters. Finally, we discuss the main results and conclusions in Sec.~\ref{sec:discussion}.

\section{\label{sec:mathematical}Mathematical background}
The equations of motion for a scalar field $\phi$ endowed with the potential $V(\phi)$, in a homogeneous and isotropic space-time with null spatial curvature, are given by
\begin{subequations}
\label{eq:2}
  \begin{eqnarray}
    H^2 &=& \frac{\kappa^2}{3} \left( \sum_j \rho_j +
      \rho_\phi \right) \, ,\quad \dot{\rho}_j = - 3 H (\rho_j + p_j ) \,
    , \\
    \dot{H} &=& - \frac{\kappa^2}{2} \left[ \sum_j (\rho_j +
      p_j ) + (\rho_\phi + p_\phi) \right] \, , \label{eq:2a} \\
    \ddot{\phi} &=& -3 H \dot{\phi} - \partial_\phi V(\phi)  \, , \label{kge}
  \end{eqnarray}
\end{subequations}
where $\kappa^2 = 8\pi G$, $\rho_j$ and $p_j$ are the energy and pressure density of ordinary matter, a dot denotes derivative with respect to cosmic time $t$, and $H = \dot{a}/a$ is the Hubble parameter. The index $j$ runs over all the matter species in the Universe apart from the scalar field (e.g.,photons, baryons, etc.). The scalar field energy density and pressure are given by the canonical expressions $\rho_\phi = (1/2)\dot{\phi}^2+V(\phi)$ and $p_\phi = (1/2)\dot{\phi}^2-V(\phi)$, whereas those of the perfect fluids are related through the barotropic relation $p_j = (\gamma_j -1) \rho_j$. The barotropic equation of state takes the usual values of $\gamma_j=4/3$ for a relativistic species and $\gamma_j=0$ for a nonrelativistic one.

We define a new set of polar coordinates in the form \cite{Copeland:1997et,Bahamonde:2017ize,Roy:2018nce,Urena-Lopez:2015gur,Urena-Lopez:2015odd,Cedeno:2017sou,doi:10.1063/1.3473869,Roy:2013wqa}, 
\begin{subequations}
\label{eq:backvars}
\begin{gather}
\frac{\kappa \dot{\phi}}{\sqrt{6} H}  \equiv  \Omega^{1/2}_\phi \sin(\theta/2), \quad \;
 \frac{\kappa V^{1/2}}{\sqrt{3} H} \equiv \Omega^{1/2}_\phi \cos(\theta/2) \, , \\
y_1  \equiv -2\ \sqrt{2} \, \frac{\partial_{\phi}V^{1/2}}{H} \, , \quad y_2 \equiv - 4\sqrt{3} \frac{\partial^2_\phi V_{\phi}^{1/2}}{\kappa H} \, ,
\end{gather}
\end{subequations}
where $\Omega_\phi = \kappa^2 \rho_\phi/(3H^2)$ is the standard density parameter associated to the quintessence density. Considering the new variables~\eqref{eq:backvars}, the Klein-Gordon equation~\eqref{kge} takes the form of the following dynamical system:
\begin{subequations}
\label{eq:new4}
  \begin{eqnarray}
  \theta^\prime &=& -3 \sin \theta + y_1 \, ,\quad  \Omega^\prime_\phi = 3 (\gamma_{tot} - \gamma_\phi)
  \Omega_\phi \label{eq:new4a} \, ,\\
  y^\prime_1 &=& \frac{3}{2}\gamma_{tot} y_1 + \Omega_{\phi}^{1/2} \sin(\theta /2) y_2 \, . \label{eq:new4b}
\end{eqnarray}
\end{subequations}
Here, a prime denotes derivative with respect to the number of $e$-foldings $N \equiv \ln (a/a_i)$, with $a$ the scale factor of the Universe and $a_i$ its initial value, and the total equation of state $\gamma_{tot} = (p_{tot} + \rho_{tot})/\rho_{tot}$. Here $p_{tot}$ ($\rho_{tot}$) denotes the total pressure (density) of all matter species under consideration in our model. Likewise, the barotropic equation of state for the quintessence field is given by $\gamma_\phi = (p_\phi + \rho_\phi)/\rho_\phi = 1 - \cos \theta$, from which we see that the standard EoS is given by $w_\phi = - \cos \theta$.

Let us now consider the case of linear perturbations $\varphi$ of the quintessence field in the form $\phi(x,t) = \phi(t)+\varphi(x,t)$. As for the metric, we choose the synchronous gauge with the line element $ds^2 = -dt^2+a^2(t)(\delta_{ij}+h_{ij})dx^idx^j$, where $h_{ij}$ is the tensor of metric perturbations. The linearized Klein-Gordon equation for a given Fourier mode $\varphi(k,t)$ reads \cite{Ratra:1990me,Ferreira:1997au,Ferreira:1997hj,Perrotta:1998vf}:
\begin{equation}
  \ddot{\varphi} = - 3H \dot{\varphi} - \left( \frac{k^2}{a^2} + \partial^2_\phi V \right) \varphi -
  \frac{1}{2} \dot{\phi} \dot{\bar{h}} \, , \label{eq:13}
\end{equation}
where a dot means derivative with respect the cosmic time, $\bar{h} = {h^j}_j$ and $k$ is a comoving wave number.

As shown in Ref.\cite{Urena-Lopez:2015gur,Cedeno:2017sou}, we can transform Eq.~\eqref{eq:13} into a dynamical system by means of the following (generalized) change of variables,
\begin{equation}
\sqrt{\frac{2}{3}} \frac{\kappa \dot{\varphi}}{H} \equiv - \Omega^{1/2}_\phi e^{\beta} \cos(\vartheta/2) \, , \; 
\frac{\kappa y_1 \varphi}{\sqrt{6}} \equiv - \Omega^{1/2}_{\phi} e^{\beta} \sin(\vartheta/2) \, , \label{eq:linearvars}
\end{equation}
with $\beta$ and $\vartheta$ the new variables needed for the evolution of the scalar field perturbations. But if we further define $\delta_0 = - e^{\beta}\sin(\theta/2-\vartheta/2)$ and $\delta_1=-e^{\beta}\cos(\theta/2-\vartheta/2)$, then Eq.~\eqref{eq:13} takes on a more manageable form,
\begin{subequations}
\label{eq:eqdeltas}
\begin{eqnarray}
\delta^\prime_0 &=&  -\left[3\sin\theta + \frac{k^2}{k^2_J}(1 - \cos \theta) \right] \delta_1 + \frac{k^2}{k^2_J} \sin \theta \delta_0 \nonumber \\
&& - \frac{\bar{h}^\prime}{2}(1-\cos\theta) \, , \label{eq:eqdeltas-a} \\
\delta^\prime_1 &=& -\left( 3\cos \theta + \frac{k^2_{eff}}{k^2_J} \sin\theta \right) \delta_1 + \frac{k^2_{eff}}{k^2_J} \left(1 + \cos \theta \right) \delta_0 \nonumber \\
&& - \frac{\bar{h}^\prime}{2} \sin \theta \, , \label{eq:eqdeltas-b}
%\delta^\prime_1 &=& -\left[3\cos \theta + \frac{k^2}{k^2_J} \sin\theta - \frac{y_2}{y_1} \Omega^{1/2}_\phi \sin (\theta/2) \right] \delta_1 \nonumber \\
%&& + \left[ \frac{k^2}{k^2_J} \left(1 + \cos \theta \right) - \frac{y_2}{y_1} \Omega^{1/2}_\phi \cos (\theta/2) \right] \, \delta_0 - \frac{\bar{h}^\prime}{2} \sin \theta \, , \label{eq:eqdeltas-b}
\end{eqnarray}
\end{subequations}
where $k_J^2 \equiv a^2 H^2 y_1$ is the (squared) Jeans wave number, a prime again denotes derivative with respect to the number of $e$-folds $N$, and 
\begin{equation}
    k^2_{eff} \equiv k^2 - \frac{y_2}{2y} a^2 H^2 \Omega_\phi \, . \label{eq:eqdeltas-c}
\end{equation}
In writing Eqs.~\eqref{eq:eqdeltas} we have used the relation $\partial^2_\phi V = H^2 (y^2_1/4 - y y_2/2)$ in Eq.~\eqref{eq:13}. 

Some notes are in turn. As shown in~\cite{Urena-Lopez:2015gur,Cedeno:2017sou}, the variable $\delta_0$ is exactly the quintessence density contrast, as one can show from Eqs.~\eqref{eq:backvars} and~\eqref{eq:linearvars} that $\delta \rho_\phi/\rho_\phi = (\dot{\phi}\dot{\varphi} + \partial_\phi V \varphi)/\rho_\phi =\delta_0$. (For a comparison of this approach to other methods in the case of a quadratic potential see\cite{Cookmeyer:2019rna}.)

The definition of the Jeans wave number $k_J$ is a generic one, and only involves the function $y_1$ (see also~\cite{Urena-Lopez:2015odd,Cedeno:2017sou} for previous applications). Actually, such definition is necessary for the transformation of Eq.~\eqref{eq:13} into the dynamical system~\eqref{eq:eqdeltas}. In the case of quintessence DE models, one expects that $y_1 \lesssim \mathcal{O}(1)$, and then the associated Jeans scale length is equal or larger than the Hubble horizon; that is, $k^{-1}_J \gtrsim 1/H$. 

Also, it must be stressed out that in general the evolution of quintessence density perturbations is driven by the effective wave number~\eqref{eq:eqdeltas-c}. A related definition of $k_{eff}$ was first presented in~\cite{Cedeno:2017sou} for an axion-like potential (for which, in our notation, $y_2 = \alpha_0 y$, see also~\cite{Urena-Lopez:2019xri}), which served to explain the tachyonic instability of scalar field perturbations whenever $k^2_{eff} < 0$. Here, Eq.~\eqref{eq:eqdeltas-c} is a generalization for any quintessence potential that indicates that tachyonic instabilities may arise whenever $y_2/y > 0$.

The inclusion of quintessence, and in general of DE, density perturbations have been neglected in most studies, mostly because it is believed that they do not have any significant influence on cosmological observables (eg~\cite{de_Putter_2010}). There are, though, recent works in the literature which are dedicated to uncover the effects that DE density perturbations can have on structure formation~\cite{Ooba:2018dzf, Mukherjee:2003rc,Zhai:2017vvt,Zhai:2017vvt,Ooba:2017lng,Park:2018fxx,Park:2019emi}. This is in agreement with our results in Sec.~\ref{sec:numerical}, where we show that linear density perturbations can help to improve constraints on quintessence models. Moreover, as we also show in Sec.~\ref{sec:without-perts}, they should necessarily be included to get correct solutions of the cosmological observables.

\section{\label{sec:critical}General tracker solutions}
The equations of motion~\eqref{eq:new4} allow us to study easily some solutions that have been considered of physical interest in the specialized literature about cosmological scalar fields. One only needs to specify the functional form of $y_2$ and then calculate the critical points of the dynamical system~\eqref{eq:new4}. For purposes of simplicity, in this work we take the parametrization of $y_2$ proposed in Ref.~\cite{Roy:2018nce},
\begin{equation} 
y_2 = y \left( \alpha_0 + \alpha_1 y_1/y + \alpha_2 y^2_1/y^2 \right) \, , \label{eq:GP1}
\end{equation}
where the $\alpha$ are just constant parameters. This parametrization includes a large class of quintessence potentials (see Table~1 and~2 in \cite{Roy:2018nce}). 

The critical condition $\theta^\prime =0$, see Eq.~\eqref{eq:new4a}, simply leads to $y_{1c} = 3 \sin \theta_c$, where a subindex $c$ denotes the critical value of the corresponding variable. Upon substitution in Eq.~\eqref{eq:new4b}, together with the general expression for $y_2$ in Eq.~\eqref{eq:GP1}, we find the remaining critical conditions,
\begin{subequations}
\label{eq:crit-phi}
\begin{align}
   \left( \gamma_{tot} + \frac{\alpha_0}{9} \Omega_{\phi c} + \frac{\sqrt{2}}{3} \alpha_1 \Omega^{1/2}_{\phi c} \gamma^{1/2}_{\phi c} + 2 \alpha_2 \gamma_{\phi c} \right) \sin \theta_c =0 \, , \label{eq:crit-phi-a} \\
   (\gamma_{tot} - \gamma_{\phi c}) \Omega_{\phi c} = 0 \, . \label{eq:crit-phi-b}
\end{align}
\end{subequations}
In writing Eq.~\eqref{eq:crit-phi-b} we used the expression $\gamma_{\phi c} = 2 \sin^2 (\theta_c/2)$ and assumed that $0 \leq \theta_c \leq \pi$. The critical solutions of Eqs.~\eqref{eq:crit-phi} are described in the Appendix~\ref{sec:critical-sols}, where we follow the original classification of Table~1 in Ref.~\cite{Copeland:1997et} (see also Table~4 in~\cite{Bahamonde:2017ize}).

Our main interest is the case of potentials with $\alpha_0=0=\alpha_1$, for which Eq.~\eqref{eq:crit-phi-a} indicates a critical condition for the quintessence EoS, 
\begin{equation}
    \gamma_{\phi c} = - \gamma_{tot}/(2\alpha_2) \, , \label{eq:tracker}
\end{equation} 
which is known as the tracker condition. Given that we expect $0 < \gamma_{\phi c} < 2$, we find that the critical condition exists for $ 0 < - \gamma_{tot}/\alpha_2 < 4$, and then $\alpha_2 <-\gamma_{tot}/4$. If the tracker condition is to be attained at very early times during radiation domination, for which $\gamma_{tot} = 4/3$, then the absolute upper bound in the active parameter is $\alpha_2 <-1/3$. 

However, one must prefer tracker solutions for which $\gamma_{\phi c} < \gamma_{tot}$, as for these instances the EoS can have better chances to approach $-1$ at late times. Thus, we will hereafter restrict our study of tracker potentials to those with $\alpha_2 < -1/2$, constraint we shall refer to as the tracker limit.

For reference, power-law quintessence $V(\phi) = M^{4-p} \phi^p$ corresponds to $p = 2/(1+2\alpha_2)$ (with $\alpha_0 = \alpha_1 =0$), and then the tracker condition~\eqref{eq:tracker} translates into $\gamma_{\phi c} = p \, \gamma_{tot}/(p-2)$. The existence condition of the tracker solution for power-law potentials is $p > 6$ (for $-1/2 < \alpha_2 < -1/3$) and $p < 0$ (for $\alpha_2 < -1/2$), whereas $\alpha_2 = -1/2$ corresponds to an exponential potential, see Table~2 in\cite{Roy:2018nce}.

Notice though that there is not in general a critical condition for the density parameter except for the trivial case $\Omega_{\phi c}=0$. Hence, the quintessence field evolves with a fixed EoS that is related to that of the dominant background component, similarly to the so-called scaling solutions (see Appendix~\ref{sec:critical-sols}), but with the difference that the quintessence density does not mimic that of the dominant component. The tracker solution is then just an approximated critical point of the quintessence equations of motion (see also the discussion in~\cite{UrenaLopez:2011ur,Gong:2014dia,Bahamonde:2017ize} about the tracker theorem in terms of dynamical systems).

However, as we have pointed out before, the tracker condition can be satisfied approximately by a more general class of potentials, even those with $\alpha_0 \neq 0$ and $\alpha_1 \neq 0$ in Eq.~\eqref{eq:GP1}, as long as the quintessence density parameter $\Omega_\phi$ is negligible with respect to the dominant one, which is regularly the case at early times in the evolution of the Universe. One can see that for practically all the quintessence potentials in Table~2 of Ref.~\cite{Roy:2018nce} it is possible to consider a tracker solution, as long as $\alpha_2 < 0$. The most known example in the literature is the inverse power-law case mentioned before, $V(\phi) \sim \phi^p$, corresponding to the so-called Class Ia, but there are others like those in Classes IIa, IIIa, and IVa that have a more involved functional form.

Moreover, all the calculations above can be extended for more general expressions of the form $y_2 = y f(y_1/y)$, where $f(y_1/y)$ is an arbitrary function of its argument~\footnote{As explained in the Appendix~A in~\cite{Roy:2018nce}, the standard roll parameter can be written as $\lambda = y_1/y$, and then $y_2/y = f(\lambda) = \lambda^2 [1-\Gamma(\lambda)]-\lambda/2$, where $\Gamma(\lambda)$ is called the tracker parameter. With this identification, one can also make a matching between the potentials of Table~1 in~\cite{Roy:2018nce} and Table~10 in~\cite{Bahamonde:2017ize}.}. The equations for the critical points for such general case are written as
\begin{subequations}
\label{eq:crit-gen}
\begin{eqnarray}
   \left[ 9 \gamma_{tot} + \Omega_{\phi c} \, f \left( \frac{3 \sqrt{2} \gamma^{1/2}_{\phi c}}{\Omega^{1/2}_{\phi c}} \right) \right] \sin \theta_c =0 \, , \label{eq:crit-gen-a} \\
   (\gamma_{tot} - \gamma_{\phi c}) \Omega_{\phi c} = 0 \, , \label{eq:crit-gen-b}
\end{eqnarray}
\end{subequations}
where we have used $y_{1c}/y_c = 3 \sin \theta_c/[\Omega^{1/2}_{\phi c} \cos(\theta_c/2)] = 3 \sqrt{2} \gamma^{1/2}_{\phi c}/\Omega^{1/2}_{\phi c}$, see also Eqs.~\eqref{eq:crit-phi}.

As before, the tracker solution is not a critical point of Eqs.~\eqref{eq:crit-gen}, but it exists whenever the quintessence density parameter can be neglected from Eq.~\eqref{eq:crit-gen-a} and the only unknown variable left in it is the quintessence EoS. That is, if the following condition is satisfied,
\begin{equation}
    \lim_{\Omega_{\phi c} \to 0} \left[ \Omega_{\phi c} \, f \left( \frac{3 \sqrt{2} \gamma^{1/2}_{\phi c}}{\Omega^{1/2}_{\phi c}} \right) \right] = g(\gamma_{\phi c}) \, , \label{eq:general-tracker}
\end{equation}
then the tracker equation derived from Eq.~\eqref{eq:crit-gen-a} simply reads: $9 \gamma_{tot} + g ( \gamma_{\phi c}) = 0$. Here, $g(\gamma_{\phi c})$ is in principle an arbitrary function of its argument obtained from the limit in Eq.~\eqref{eq:general-tracker}. From the foregoing equation we find the tracker value of the quintessence EoS, if such solution exists and satisfies the constraint $0 < \gamma_{\phi c} < 2$.

One example is the potential $V = M^4 e^{1/\kappa \phi}$\cite{Steinhardt:1999nw}, for which $y_2= -y ( 6^{3/4} y^{3/2}_1/y^{3/2} + y^2_1/y^2)$. We can see that this potential is not covered by the polynomial relation~\eqref{eq:GP1}, but it complies with the condition~\eqref{eq:general-tracker}. From the latter we obtain that $g(\gamma_{\phi c}) = -18 \gamma_{\phi c}$, and the corresponding tracker solution at early times simply reads $\gamma_{\phi c} = \gamma_{tot}/2$.

\section{\label{sec:numerical}Numerical solutions and comparison with observations}
One critical step in the numerical solution of Eqs.~\eqref{eq:2} and~\eqref{eq:new4} is to find the correct initial conditions of the dynamical variables. There is not a general recipe for quintessence fields, and one must choose expressions case by case. However, the tracker condition~\eqref{eq:tracker} simplifies the numerical effort, and for the initial conditions of the dynamical variables we obtain
\begin{subequations}
\label{eq:initial}
\begin{eqnarray}
\cos \theta_i &=& 1 + \frac{2}{3\alpha_2} \, \quad y_{1i} = 3 \sin \theta_i \, , \label{eq:initial-a} \\ 
\Omega_{\phi i} &=& A \times a^{4(1+1/2\alpha_2)}_i \left( \frac{\Omega_{m 0}}{\Omega_{r 0}} \right)^{1+1/2\alpha_2} \Omega_{\phi 0} \, , \label{eq:initial-b}
\end{eqnarray}
\end{subequations}
where $\Omega_{r0}$, $\Omega_{m0}$ and $\Omega_{\phi 0}$ are, respectively, the present density parameters of relativistic matter, nonrelativistic matter and quintessence, and $a_i$ is the initial value of the scale factor (typically $a_i \simeq 10^{-14}$). The expressions for $\theta_i$ and $y_{1i}$ correspond to the tracker condition~\eqref{eq:tracker}, whereas the one for $\Omega_{\phi i}$ is derived from integrating Eq.~\eqref{eq:new4a} up to present time assuming the tracker evolution during radiation and matter domination eras, with $A$ an arbitrary constant.
 
We have verified that Eqs.~\eqref{eq:initial} provide good enough initial values for numerical solutions in the general case\footnote{As mentioned before, see Eq.~\eqref{eq:general-tracker} and below it, the values $(\alpha_0=0 = \alpha_1,\alpha_2=-1/2)$ correspond to an exponential potential for the chosen parametrization~\eqref{eq:GP1}. For this case the tracker critical solution~\eqref{eq:tracker} gives $\gamma_{\phi c} = \gamma_{tot}$, which is in fact the typical scaling solution of exponential potentials~\cite{Copeland:1997et,Ferreira:1997au,Ferreira:1997hj}. Hence, the tracker limit $\alpha_2 < -1/2$ also helps us to avoid any overlapping between tracker and scaling solutions.}. This is especially important as we rely on an amended version of the the Boltzmann code \texttt{CLASS} (v2.5)~\cite{Lesgourgues:2011rg,*Blas:2011rf,*Lesgourgues:2011re,*Lesgourgues:2011rh} to adjust the value of the coefficient $A$, so that we obtain the correct value of the quintessence density parameter $\Omega_{\phi 0}$ at the present time. For the initial conditions of the linear perturbations we simply use $\delta_{0i} = 0$ and $\delta_{1i} = 0$, as the evolution of the perturbation variables is mostly driven by the nonhomogeneous terms in Eqs.~\eqref{eq:eqdeltas}.

\begin{figure}[htp!]
\includegraphics[width=0.43\textwidth]{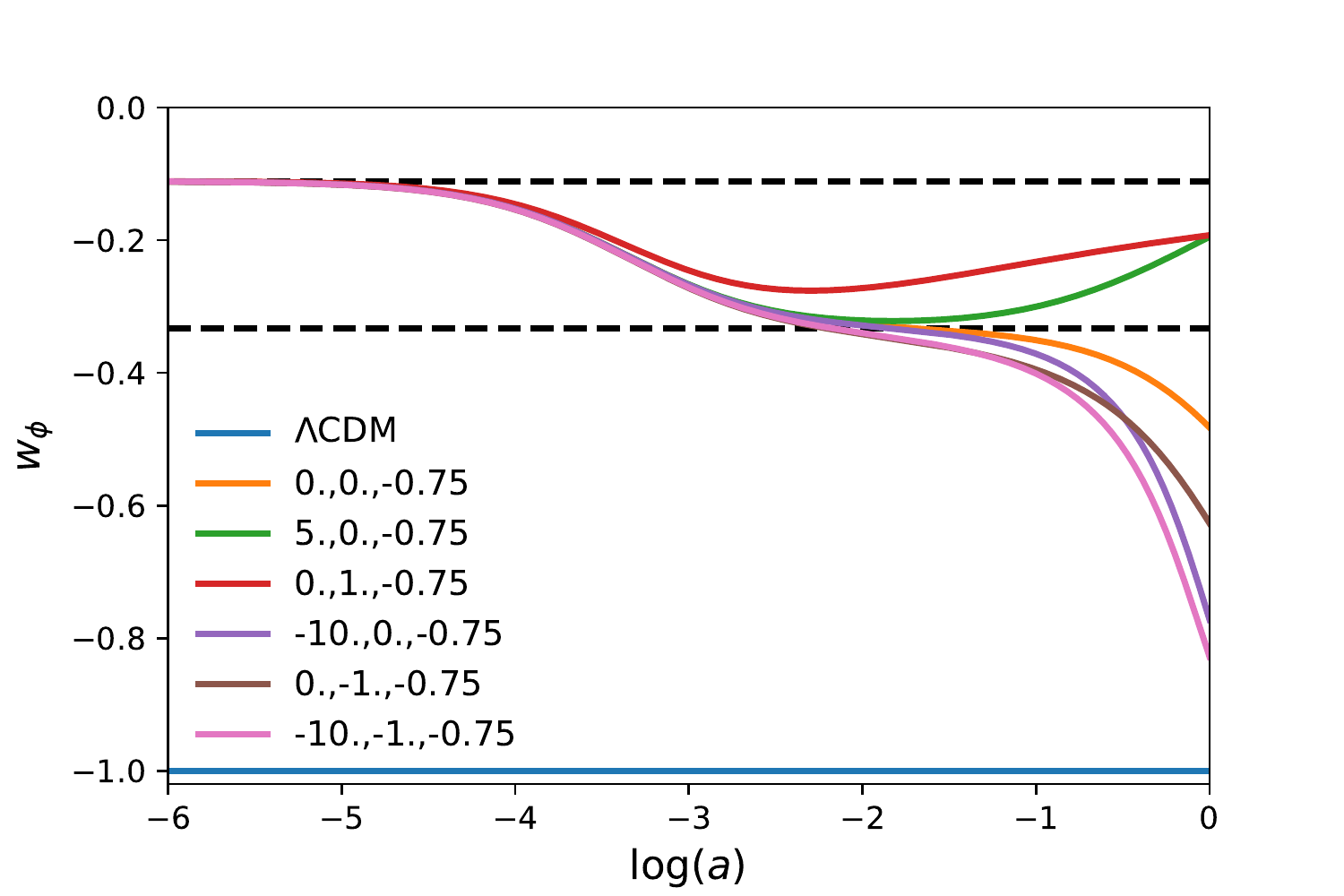}
\includegraphics[width=0.43\textwidth]{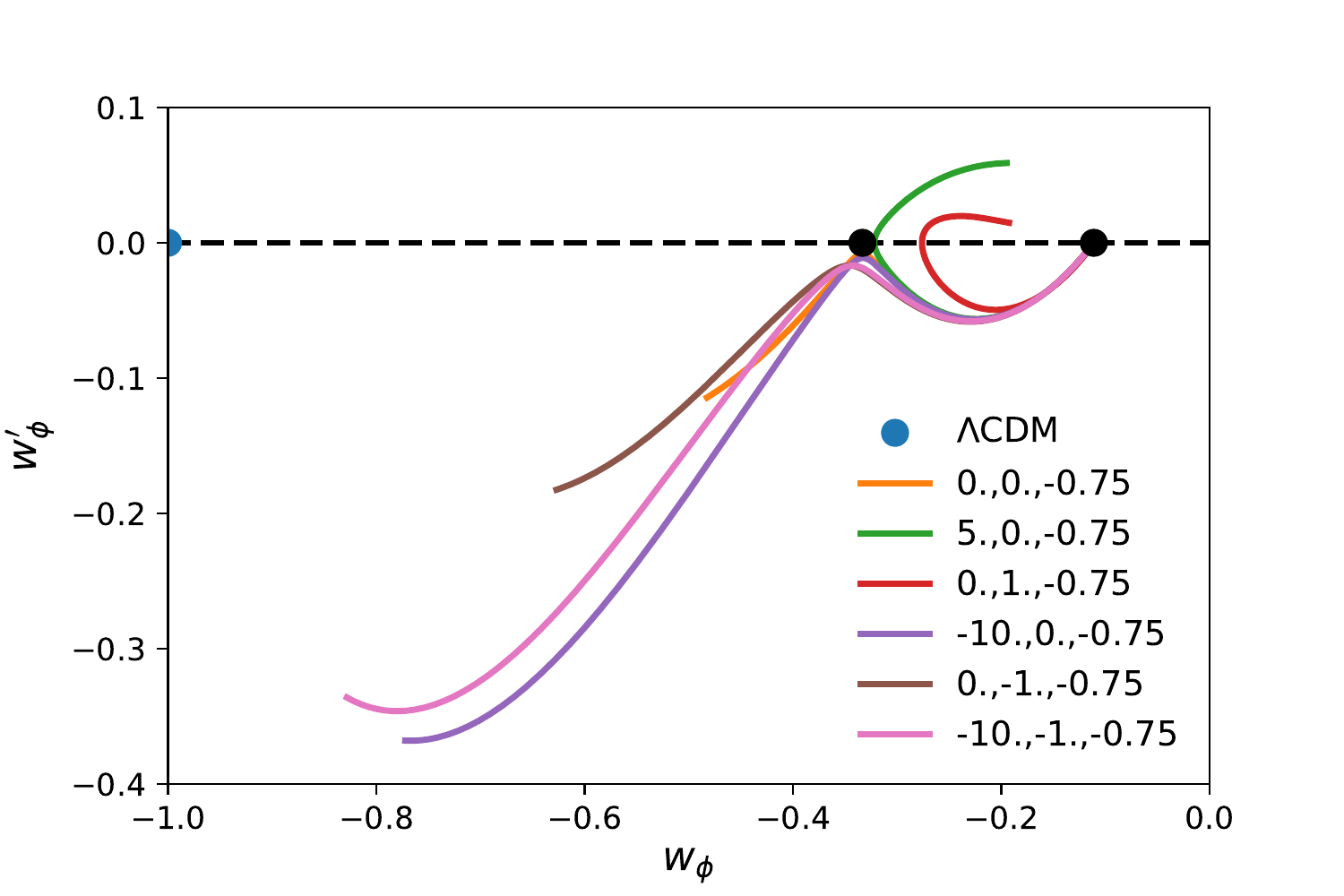}

\caption{\label{fig:numerics-a} (Top) The evolution of the EoS $w_\phi$ for tracker quintessence models with $\alpha_2 =-3/4$, see Eq.~\eqref{eq:GP1}. Each curve represents a model with the indicated values of the triplet $(\alpha_0,\alpha_1,\alpha_2)$. The first example in all figures corresponds to the inverse power-law potential $V(\phi)=M^8 \phi^{-4}$ ($\alpha_0=0=\alpha_1$). The other cases that additionally consider ($\alpha_0 \neq 0 \neq \alpha_1$) can be identified by their corresponding color. The case of $\Lambda$CDM is also shown for reference. The dashed horizontal lines represent the tracker values of the quintessence EoS: $w_\phi = -1/9$ (radiation domination) and $w_\phi = -1/3$ (matter domination). (Bottom) Phase space of the EoS $(w_\phi,w^\prime_\phi)$, for the same cases (with the same colors) as in the top panel. The blue dot represents the cosmological constant case, whereas the black dots represent the aforementioned tracker values at $(-1/9,0)$ and $(-1/3,0)$. See the text for more details.}
\end{figure}

For purposes of illustration, in Figs.~\ref{fig:numerics-a} we show some numerical examples for the case $\alpha_2=-3/4$, whereas other parameters, like the present density contributions of the different matter species, were fixed to the values reported by the Planck collaboration (see their Table~1)\cite{Aghanim:2018eyx}. 

As explained in Ref.~\cite{Roy:2018nce}, $\alpha_2=-3/4$ (with $\alpha_0 =0 = \alpha_1$) corresponds to the power-law potential $V(\phi)= M^8 \phi^{-4}$. In the top panel of Fig.~\ref{fig:numerics-a} we show the influence of the other two active parameters $\alpha_0$ and $\alpha_1$ in the final evolution of the quintessence EoS. Notice that the original tracker solution reaches its tracker values at early times (represented by the dashed black lines), but it cannot accelerate the expansion of the Universe at late times (as one requires $w_\phi < -2/3$). The latter flaw can be corrected by considering potentials with power in the range $-4 < p < 0$\cite{Ooba:2018dzf}, but also by considering negative values of the other active parameters. The more negative the latter are, the more the quintessence EoS gets closer to the cosmological constant case $w_\phi \to -1$.

Another view of the evolution of the EoS is shown in the bottom panel of Fig.~\ref{fig:numerics-a}, in terms of the phase space $(w_\phi,w^\prime_\phi)$, where the derivative of the EoS is calculated from $w^\prime_\phi = -\sin \theta (3 \sin\theta -y_1)$. It can be seen that all solutions depart from the tracker solution at radiation domination $(-1/9,0)$, and then evolve toward that at matter domination $(-1/3,0)$. However, for the cases in which $\alpha_0,\alpha_1 \leq 0$, the curves are deflected away from the second tracker point and the EoS evolves toward the cosmological constant point at $(-1,0)$. Likewise, the cases with $\alpha_0,\alpha_1 > 0$ are deflected in the opposite direction, the derivative is positive and then the EoS evolves toward less negative values.

The behavior of the EoS can be understood in terms of the critical point that corresponds to quintessence domination, which is described in the Appendix~\ref{sec:critical-sols}. As explained there, two critical points of the EoS always exist, which correspond to the following points in the phase space: $(1,0)$ and $(-1,0)$. The first one is known in the literature as the kinetic domination critical point, whereas the second is the cosmological constant case. 

But there are other possibilities if the active parameters take on appropriate values. It can be shown that for the type of examples in Fig.~\ref{fig:numerics-a}, for which $\alpha_2 < -1/2$, we also find that
\begin{subequations}
\label{eq:critw}
\begin{eqnarray}
    1 + w_{\phi c} &=& - \frac{\alpha_0}{9(1+2\alpha_2)} \, , \quad \alpha_0 > 0 \, , \alpha_1 = 0 \, , \label{eq:critw-a} \\
   &=& \frac{2 \alpha^2_1}{9(1+2\alpha_2)^2} \, , \quad \alpha_0 = 0 \, , \alpha_1 > 0 \, . \label{eq:critw-b}
\end{eqnarray}
\end{subequations}
Even though the second solution~\eqref{eq:critw-b} involves $\alpha^2_1$, it must be noticed that the corresponding solution of Eq.~\eqref{eq:quint-dom} necessarily requires $\alpha_1 > 0$. Given that at the critical points $y_{1c} = 3 \sin \theta_c$, we also obtain that $w^\prime_{\phi c} =0$.

For the particular cases shown in Fig.~\ref{fig:numerics-a} with $\alpha_2 = -3/4$, we find the following critical points in the phase space: $(1/9,0)$ if $\alpha_0 = 5$ and $\alpha_1=0$ (green curves), and $(-1/9,0)$ if $\alpha_0 = 0$ and $\alpha_1=1$ (red curves). The curves in the bottom panel of Fig.~\ref{fig:numerics-a} seem to be in agreement with these calculations, although we would have to evolve further the numerical solutions, until full quintessence domination is reached, to eventually see the curves approaching the foregoing critical points. 

There are other, more general, solutions of Eq.~\eqref{eq:quint-dom} for which $\alpha_0$ and $\alpha_1$ are both positive, but they will correspond, if existent, to values of the EoS larger than the tracker solution during matter domination, for which the quintessence component will also be unable to accelerate the expansion of the Universe.

\begin{figure}[htp!]
\includegraphics[width=0.43\textwidth]{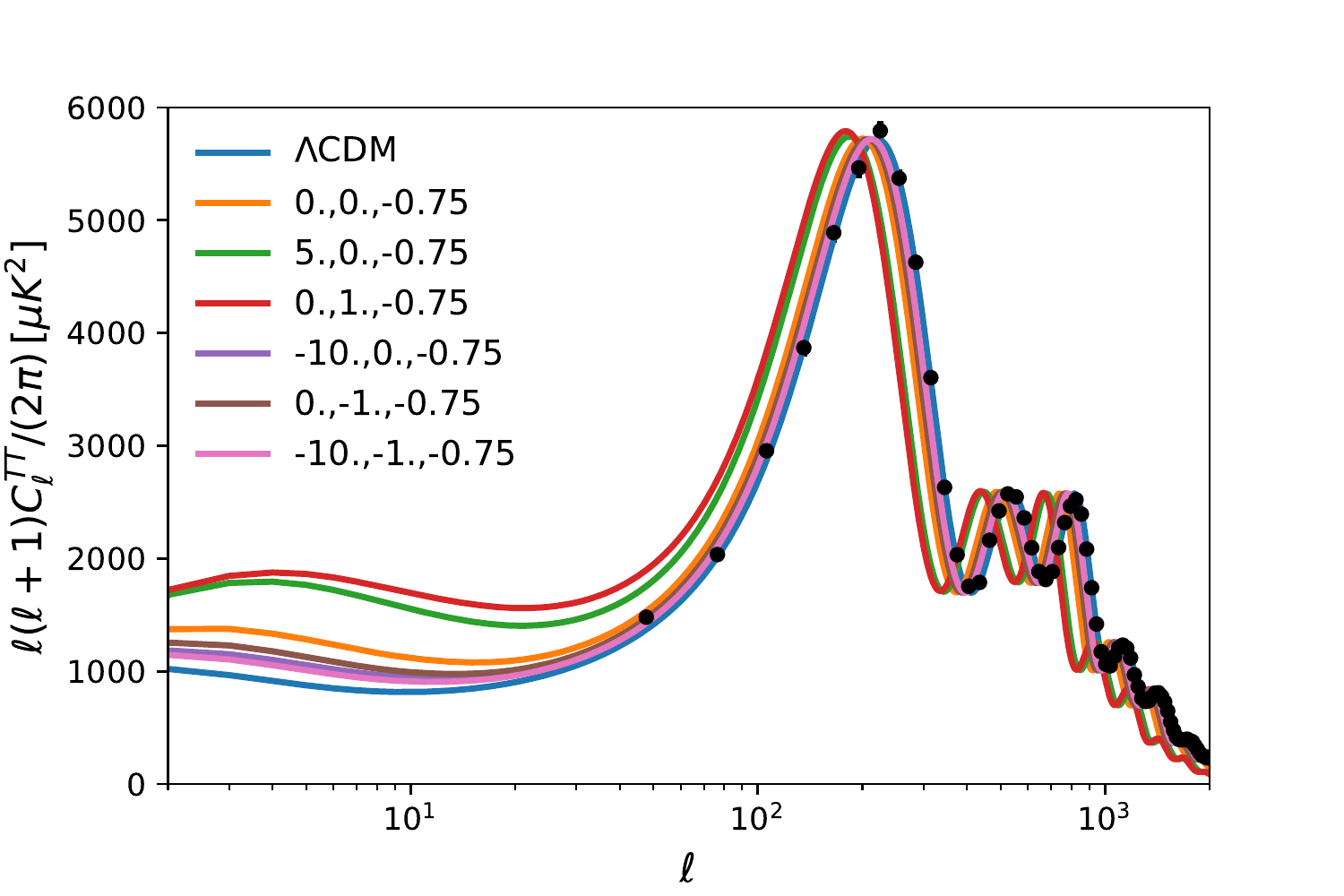}
\includegraphics[width=0.43\textwidth]{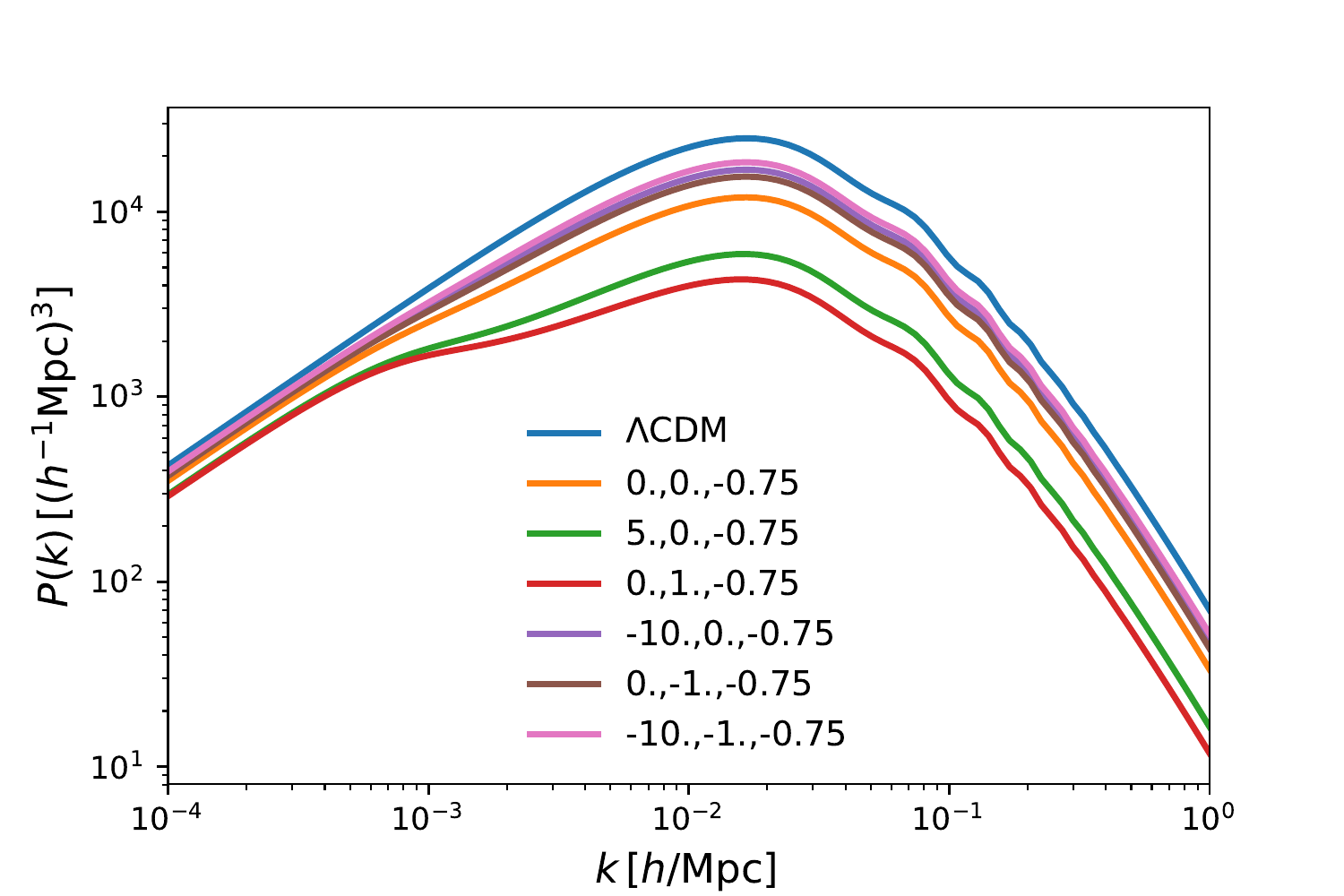}
\caption{\label{fig:numerics-b} The anisotropies of the CMB (top panel) and the MPS (bottom panel) for the same models shown in Fig.~\ref{fig:numerics-a}. The case of $\Lambda$CDM is also shown for reference in each case. The dots in the top panel are the binned TT power spectrum from the Planck collaboration. See the text for more details.}
\end{figure}

We show in Fig.~\ref{fig:numerics-b} the two-point temperature autocorrelation power spectrum $C^{TT}_\ell$ of the cosmic microwave background (CMB) and the mass power spectrum (MPS) of linear density perturbations $P(k)$, for the same numerical examples shown in Fig.~\ref{fig:numerics-a}. Notice that all the tracker cases can be easily distinguished from the $\Lambda$CDM case, and in general there are changes in the amplitude and location of the characteristic features of the observables. 

The cases in which the EoS deviates the most from the accelerating regime also show the major differences in the CMB anisotropies and the MPS with respect to the $\Lambda$CDM results, and then the former can be a strong tool to constrain the tracker models. Actually, one again sees that the active parameters $\alpha_0$ and $\alpha_1$ help to close the gaps with respect to $\Lambda$CDM if they take on negative values. We must stress out that the numerical results in Fig.~\ref{fig:numerics-b} include the contribution of quintessence density perturbations, as described in Sec.~\ref{sec:critical} above, as otherwise one obtains misleading outputs for the CMB anisotropies and MPS, as we explain in Appendix~\ref{sec:without-perts}.

We use the aforementioned amended version of the Boltzmann code
\texttt{CLASS} and the Monte Carlo code \texttt{MONTE PYTHON} (v3.2)~\cite{Brinckmann:2018cvx,*Audren:2012wb}. We consider two data sets that are sensitive to the background dynamics: (i) the Pantheon supernova data\cite{Scolnic:2017caz}, (ii) BAO (baryonic acoustic oscillations) measurements\cite{Alam:2016hwk,Beutler:2011hx,Ross:2014qpa,Agathe:2019vsu,Blomqvist:2019rah}, together with a Planck2018\cite{Aghanim:2018eyx} prior on the baryonic matter component: $\omega_b = 0.02230 \pm 0.00014$. 

The total set of parameters being sampled are: the active parameters $\alpha$, the baryonic and dark matter components, $\omega_b$ and $\omega_{cdm}$, respectively, and the supernovae nuisance parameter $M$, whereas $\Omega_{\phi 0}$ is set by the Friedmann closure relation. The set of derived parameters is: the density parameters of total matter $\Omega_M$ and quintessence $\Omega_\phi$, and the quintessence EoS $w_\phi$.

\begin{figure}[htp!]
\includegraphics[width=0.43\textwidth]{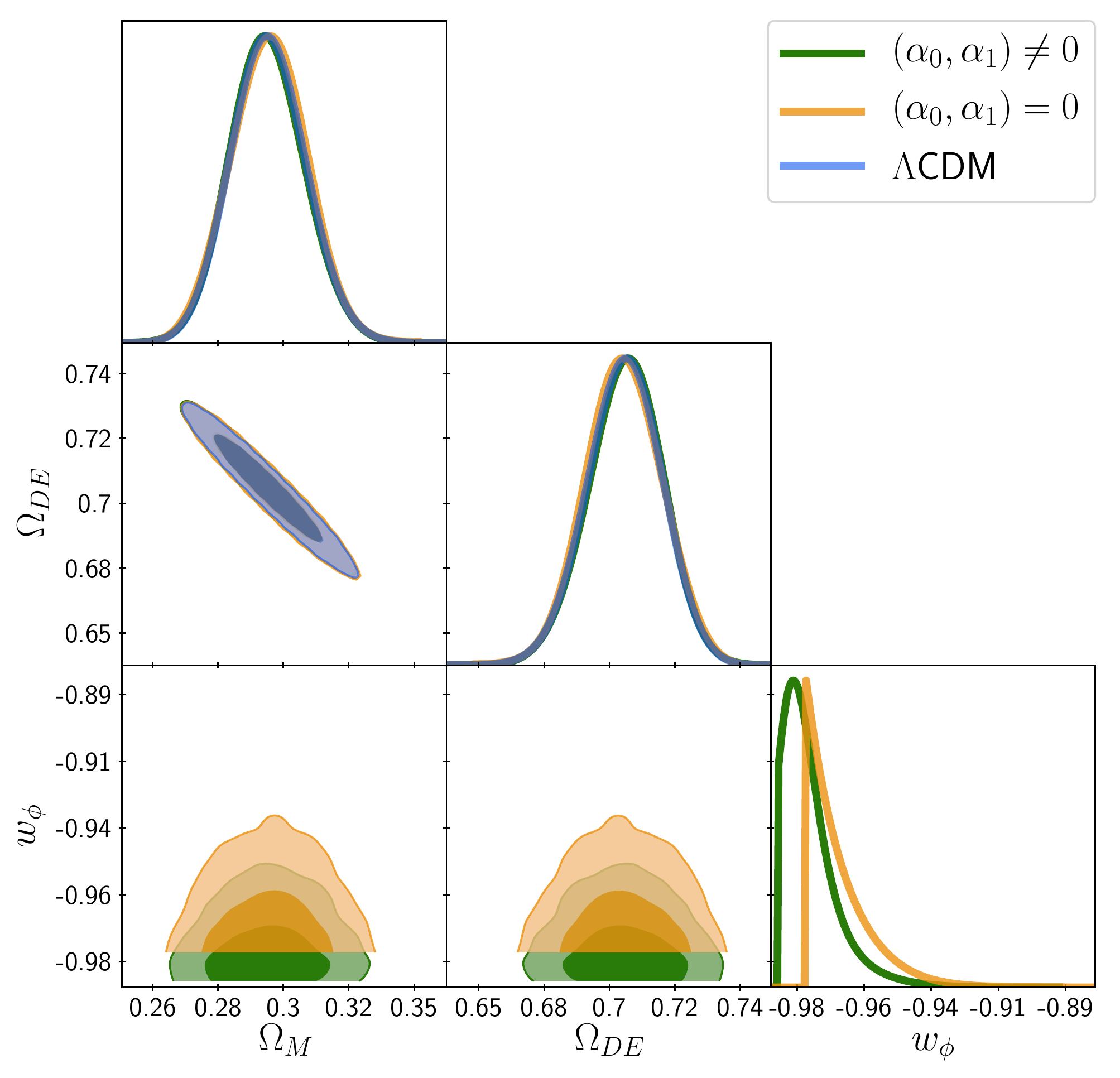}
\includegraphics[width=0.43\textwidth]{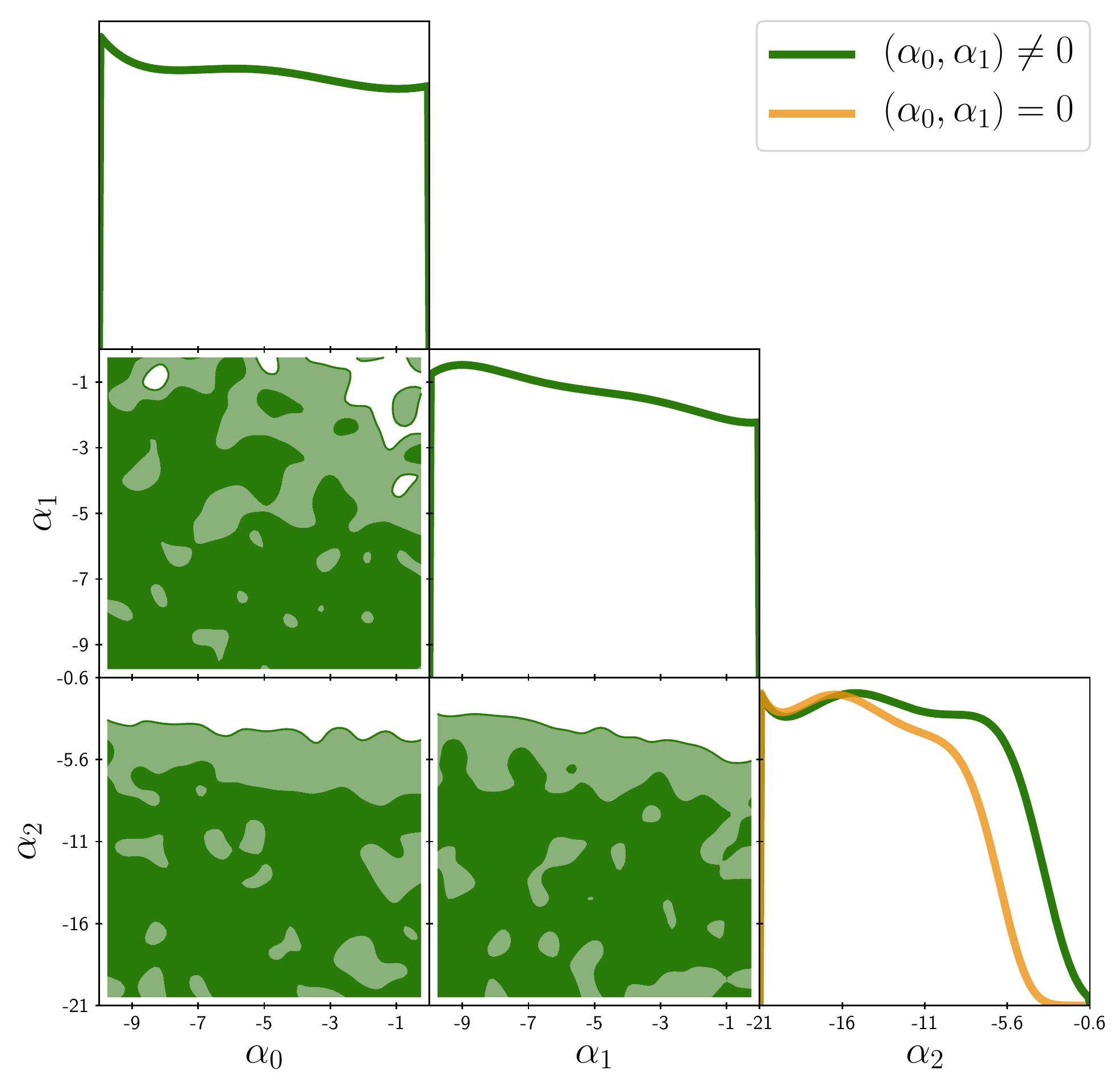}
\caption{\label{fig:quadratic} Observational constraints on $\Omega_{DE}$, $\Omega_m$, $w_\phi$ (top) and the active parameters $\alpha_0$, $\alpha_1$ y $\alpha_2$ (bottom) for the same type of tracker potentials shown in Figs.~\ref{fig:numerics-a} and~\ref{fig:numerics-b}. Here $\Omega_{DE}$ denotes the contributions of the DE component, whether the quintessence one or the cosmological constant in the $\Lambda$CDM model. See the text for more details.}
\end{figure}

As an abridged version of the numerical results, we show in the top panel of Fig.~\ref{fig:quadratic} the confidence regions for $\Omega_m$, the quintessence EoS $w_\phi$ and $\Omega_\phi$, for tracker potentials, with and without the participation of the active parameters $\alpha_0$ and $\alpha_1$. It can be seen that the quintessence EoS is closer to $-1$ if the other active parameters are included, and then the resultant physical quantities are in turn closer to those of the $\Lambda$CDM model. For the case $(\alpha_0,\alpha_1) =0$ we obtain $w_\phi < -0.948$ ($2\sigma$), whereas for $(\alpha_0,\alpha_1) \neq 0$ the result is $w_\phi < -0.961$ ($2\sigma$).

Likewise, we see in the bottom panel of Fig.~\ref{fig:quadratic} that the active parameters $\alpha_0$ and $\alpha_1$ appear unconstrained for the range they were sampled $[-10:0]$, whereas $\alpha_2$ can even take values closer to the tracker limit $\alpha_2 \simeq -1/2$. More precisely, for $(\alpha_0,\alpha_1) =0$ the constraint is $\alpha_2 < -6.63$ ($2\sigma$), whereas for $(\alpha_0,\alpha_1) \neq 0$ the result is $\alpha_2 < -4.13$ ($2\sigma$).

\section{\label{sec:discussion}Discussion and conclusions}
We have revised the case of tracker quintessence models of DE using the same formalism and parametrization presented before in Refs.~\cite{Roy:2018nce}, but now also including the influence of linear perturbations of the quintessence field following the prescription in~\cite{Urena-Lopez:2019xri,Cedeno:2017sou,Urena-Lopez:2015odd,Urena-Lopez:2015gur}. As discussed above, the new formalism allowed the identification of new tracker models which are able to have a behavior more similar to that of the cosmological constant of $\Lambda$CDM. Moreover, apart from the standard background dynamics, we were even able to include linear density perturbations of the quintessence field within the same parametrization scheme of the potentials.

We identified the necessary condition to be satisfied if a quintessence potential is to have a tracker behavior; interestingly enough, the condition involves just one of the so-called active parameters in the parametrization of the potential. We showed that such condition applies even in the presence of the other active parameters, which guarantees the tracker behavior and in turn ameliorates the fine tuning of the initial conditions at early times; in fact, we found an analytical form of the tracker initial conditions that was successfully used in our numerical studies of the models.

The other active parameters play a role in the late time dynamics of the quintessence field, and we showed that the latter behaves more similar to the cosmological constant if they take on negative values. This means that tracker quintessence models with an acceptable late-time dynamics can have a more involved functional form, for instance beyond the ubiquitous inverse-power-law one, than of those considered in the literature, which were usually chosen mostly by its early tracker behavior only.

Another advantage of our formalism is that linear density perturbations can be easily included in the numerical calculations, and then one can study the influence on them of the active parameters in the quintessence potential. Even though the common wisdom is to neglect DE density perturbations, we showed that they should be included for full consistency in the calculation of observables. In the particular case of tracker potentials, it was clear that density perturbations are important also to have more stringent constraints on the parameters of the models.

One common criticism of quintessence models is that one can always obtain from them an accelerating solution by means of a proper fine-tuning of the free parameters in the potential~\cite{Caldwell_2005,Escamilla_Rivera_2016}, which may also be applied to our parametrization. However, one key difference is that our formalism allows the identification of the tracker condition, with its corresponding attractor behavior, without spoiling the accelerating dynamics at late times. Another key difference is that we are including density perturbations in a consistent manner, which also helps to break the degeneracy between generic DE models and quintessence. It is also worth mentioning here that the contribution of the quintessence field at the early times must be negligible for the quintessence field to track the background. Hence, for this kind of tracker models early dark energy~\cite{Poulin_2019}, or any similar variant of an early contribution of the quintessence field, may not be realizable.

Although the parametrization chosen is suitable for some classes of potentials, it can be amended and extended to include other more non-conventional quintessence models. Such study, and the corresponding observational consequences derived from it, will be presented elsewhere.

\begin{acknowledgments}
This work was partially supported by Programa para el Desarrollo Profesional Docente; Direcci\'on de Apoyo a la Investigaci\'on y al Posgrado, Universidad de Guanajuato, research Grant No. 036/2020; CONACyT M\'exico under Grants No. A1-S-17899, 286897, 297771; and the Instituto Avanzado de Cosmolog\'ia Collaboration.
\end{acknowledgments}

\appendix

\section{CRITICAL POINTS OF QUINTESSENCE EQUATION OF MOTION \label{sec:critical-sols}}
Here we briefly describe the critical solutions obtained from Eq.~\eqref{eq:crit-phi} for potentials that can be written in the form~\eqref{eq:GP1}, as said before we follow here the classification already known in the literature.

\paragraph{Fluid-dominated solution.} This corresponds to the condition $\Omega_{\phi c}=0$ in Eq.~\eqref{eq:crit-phi-b}, and then the contribution of the quintessence field to the cosmic density is null. Additionally, Eq.~\eqref{eq:crit-phi-a} indicates that necessarily $\theta= 0,\pi$, and then $\gamma_c =0,2$.

\paragraph{Scaling solution.} The other possibility from Eq.~\eqref{eq:crit-phi-b} is $\gamma_{tot} = \gamma_c$, so that the quintessence field in this solution has the same EoS as the dominant background component. Moreover, now Eq.~\eqref{eq:crit-phi-b} gives the general condition,
\begin{equation}
    \alpha_0 \Omega_{\phi c} + 3\sqrt{2} \alpha_1 \, \Omega^{1/2}_{\phi c} \, \gamma^{1/2}_{tot} + 9 (1+2 \alpha_2) \gamma_{tot} =0 \, . \label{eq:quint-scaling}
\end{equation}

A scaling solution of $\Omega_{\phi c}$ from Eq.~\eqref{eq:quint-scaling} will only exist for certain combinations of the active parameters $\alpha$ so that the term inside the square brackets is positive definite and $0 < \Omega_{\phi c} < 1$. For instance, it can be verified that in the case $\alpha_1 =0=\alpha_2$ we recover the standard expression of the scaling solution reported in the literature: $\Omega_{\phi c} = -9\gamma_{tot}/\alpha_0$, which only exists for potentials with $\alpha_0 < 0$\cite{Urena-Lopez:2019xri,Matos:2000ss,*Matos:2000ng,Sahni:1999qe,Copeland:1997et,Bahamonde:2017ize,Urena_Lopez_2019}. However, Eq.~\eqref{eq:quint-scaling} opens the possibility for a generalized scaling solution, similarly to the generalized tracker quintessence discussed in the main text, for potentials with nonzero active parameters.

\paragraph{Quintessence-dominated solution.} This solution is characterized by the conditions $\Omega_{\phi c} = 1$ and $\gamma_{tot} = \gamma_{\phi c}$, under which Eq.~\eqref{eq:crit-phi-a} now reads
\begin{equation}
     \left[ \alpha_0 + 3\sqrt{2} \alpha_1 \, \gamma^{1/2}_{\phi c} + 9 (1+2 \alpha_2) \gamma_{\phi c} \right] \sin \theta_c =0 \, . \label{eq:quint-dom}
\end{equation}
The simplest possibilities are $\theta_c = 0,\pi$, for which the quintessence EoS is $\gamma_{\phi c}=0,2$, respectively. This means that the quintessence density is dominated either by its potential ($V(\phi)$) or kinetic ($\frac{1}{2}\dot{\phi}^2$) part, respectively too. Other critical solutions of the EoS can arise from the part of Eq.~\eqref{eq:quint-dom} inside the square brackets. They will depend on the particular values of the active parameters $\alpha$, as long as $0 < \gamma_{\phi c} < 2$, in which case one obtains a quintessence dominated stage with a well-defined value of the EoS that depends on the active parameters $\alpha$. 

\section{TRACKER QUINTESSENCE WITHOUT LINEAR DENSITY PERTURBATIONS \label{sec:without-perts}}
\begin{figure*}[tp!]
\includegraphics[width=0.43\textwidth]{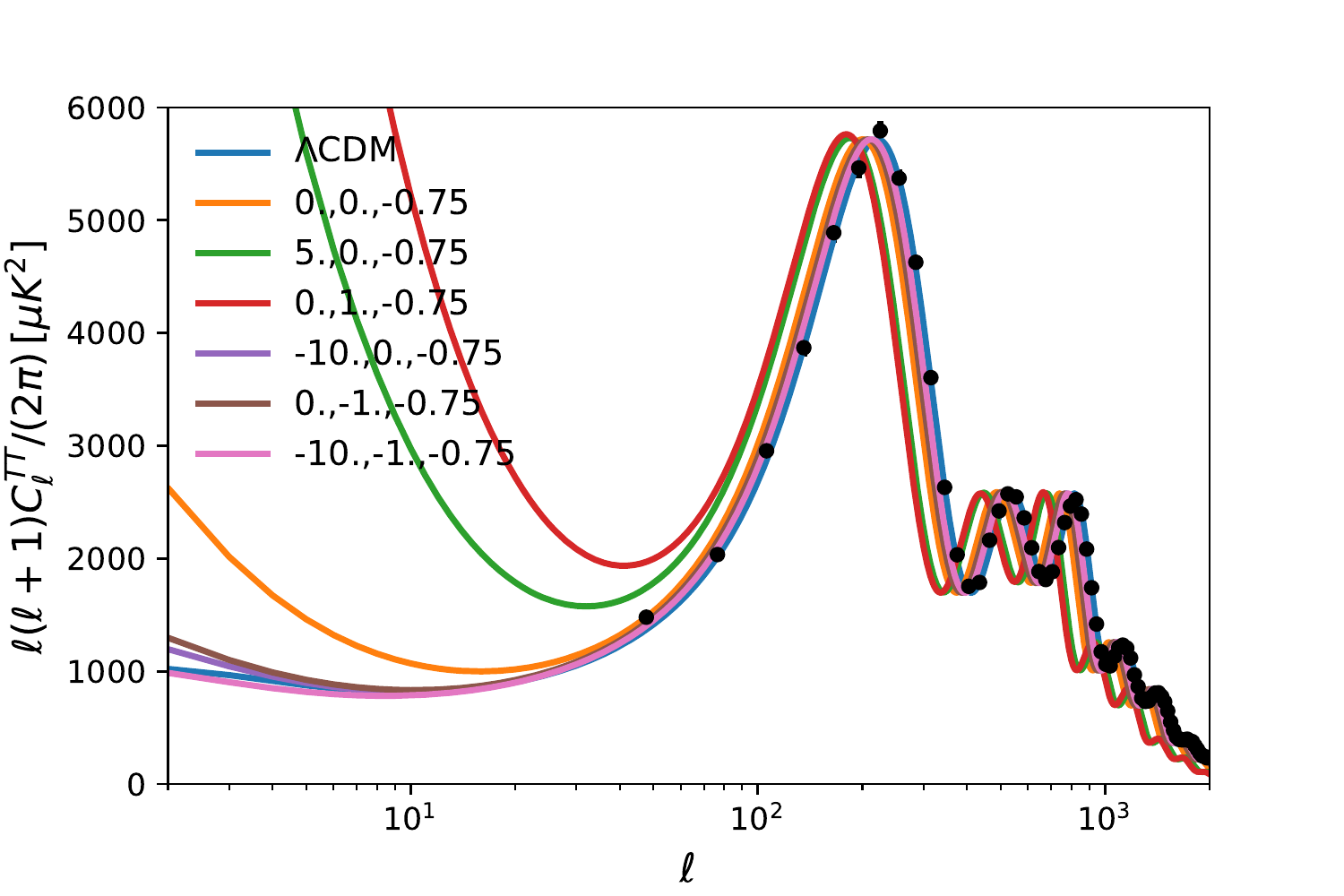}
\includegraphics[width=0.43\textwidth]{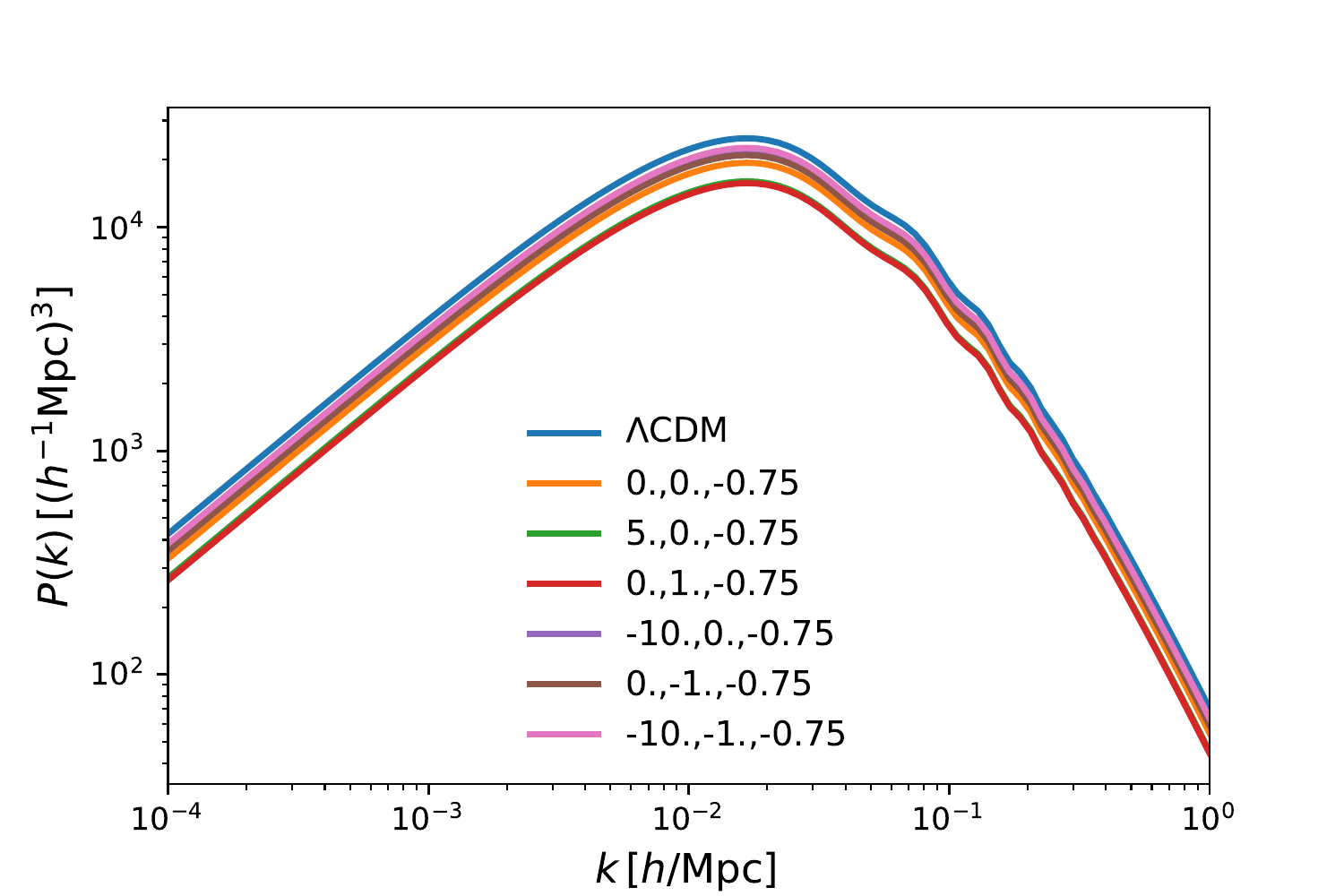}
\caption{\label{fig:numerics-x} The same tracker quintessence models as in Fig.~\ref{fig:numerics-b}, but without considering linear density perturbations for the quintessence field. See the text for more details.}
\end{figure*}

To highlight the importance of linear density perturbations in DE studies, we show in Fig.~\ref{fig:numerics-x} the same cosmological quantities as in Fig.~\ref{fig:numerics-b}, for the same tracker quintessence models, but without the linear density perturbations from the quintessence field. 

A quick comparison between Figs.~\ref{fig:numerics-b} and~\ref{fig:numerics-x} show that one can obtain misleading constraints on the quintessence models if density perturbations are neglected. In particular, we notice that the CMB power spectrum is miscalculated for some choices of the tracker models, specially at low multipoles. As for the MPS, without quintessence perturbations it may seem that the results are similar to those of $\Lambda$CDM, but this again would be misleading: from Fig.~\ref{fig:numerics-b} we clearly learn that quintessence tracker models can have a heavy influence on the evolution of the MPS, and the latter indeed becomes a useful tool to discriminate between different types of tracker quintessence models.

%\nocite{*}

\bibliography{apssamp}% Produces the bibliography via BibTeX.

\end{document}